\documentstyle[aps,epsf]{revtex}

\newcommand{\beq}{\begin{equation}}
\newcommand{\eeq}{\end{equation}}
\newcommand{\beqa}{\begin{eqnarray}}
\newcommand{\eeqa}{\end{eqnarray}}

\def\IRAS{{\it IRAS\/} }

\def\lexp{\langle}
\def\rexp{\rangle}
\def\dD{\delta_{\rm D}} 
\def\dt{\tilde \delta_m}

\def\aap{Astron. Ap. }

\def\apj{Astrophys. J. }

\def\mnras{Mon. Not. R. Astron. Soc. }

\def\k{{\bbox{k}}}
\def\x{{\bbox{x}}} 

\def\max{{\rm max}}
\def\Jy{\, {\rm Jy}}
\def\Mpc{\, h^{-1} \, {\rm Mpc}}
\def\kMpc{\, h \, {\rm Mpc}^{-1}}
\def\sk{\noalign{\medskip}}

\begin{document}

\draft

\wideabs%
{

\title{Constraints on Galaxy Bias, Matter Density, and Primordial
Non--Gausianity from the PSC$z$ Galaxy Redshift Survey}

\author{
Hume A. Feldman,$^{1}$  Joshua A. Frieman,$^{2,3}
$ J. N. Fry,$^{4}$   and Rom\'{a}n Scoccimarro$^{5}$
}

\address{
$^1$Department of Physics \& Astronomy, University of Kansas, 
Lawrence KS 66045 feldman@ukans.edu \\
$^2$NASA/Fermilab Astrophysics Center, Fermi National 
Accelerator Laboratory, Batavia, IL  60510 frieman@fnal.gov\\
$^3$Department of Astronomy and Astrophysics, University 
of Chicago, Chicago, IL 60637 \\
$^4$Department of Physics, University of Florida,
Gainesville FL 32611-8440 fry@phys.ufl.edu\\
$^5$Institute for Advanced Study, School of Natural Sciences,
Einstein Drive, Princeton, NJ 08540 scoccima@ias.edu
}

\maketitle

\begin{abstract}

We compute the bispectrum for the \IRAS PSC$z$ catalog and find that the
galaxy distribution displays the characteristic signature of
gravity. Assuming Gaussian initial conditions, we obtain galaxy biasing
parameters $1/b_1=1.20^{+0.18}_{-0.19}$ and $b_2/b_1^2=-0.42\pm0.19$,
with no sign of scale-dependent bias for $k\leq 0.3$ h/Mpc. These
results impose stringent constraints on non-Gaussian initial conditions.
For dimensional scaling models with $\chi^2_N$ statistics, we find
$N>49$, which implies a constraint on primordial skewness $B_3<0.35$.

\end{abstract}


}

{\it Introduction.}---What is the nature of the distribution of matter
in the universe?  Does the galaxy distribution follow the mass
distribution on large scales? Our ability to constrain cosmological
parameters from galaxy clustering is only as good as our understanding
of bias, that is, the relation between the galaxy distribution and the
underlying dark matter distribution.  In the current structure formation
paradigm, most of the matter in the universe is dark, interacting only
through gravity, and cold.  Small fluctuations, generated in the early
universe, grow gravitationally once the universe becomes
matter-dominated, leading to the formation of large-scale structure.
Non-linear gravitational instability (GI) theory makes specific
predictions for the statistical properties of the mass distribution,
e.g., how non-Gaussian features arise from Gaussian initial conditions.
The extent to which the galaxy distribution follows these predictions
can be used to put constraints upon galaxy bias (and thus upon the
physics of galaxy formation) and upon the degree of non-Gaussianity of
the primordial fluctuations (which constrains the physics of the
generation of fluctuations in the early universe).

In this Letter, we report and interpret measurements of the bispectrum
for the \IRAS PSC$z$ galaxy redshift survey \cite{PSCz}.  Details of the
method and earlier results can be found in two previous papers.  In the
first~\cite{PI}, we developed a framework to obtain reliable constraints
upon galaxy biasing and primordial non-Gaussianity that takes into
account the effects of survey geometry and sampling.  In the
second~\cite{PII}, we computed the bispectrum of the galaxy distribution
of earlier \IRAS redshift catalogs.  Due to its larger volume and higher
sampling density, the PSC$z$ survey yields the best evidence supporting
GI and the strongest constraints to date upon galaxy biasing and
primordial non-Gaussianity.

{\it The Bispectrum.}---In linear perturbation theory (PT), the
statistical properties of the primordial fluctuations, Gaussian or
non-Gaussian, are preserved.  If Gaussian, they can be fully
characterized in Fourier space by their power spectrum $P(k)$,
\beq
\lexp \dt(\k_1) \dt(\k_2) \rexp = (2\pi)^3 \dD(\k_1+\k_2) \, P_m(k_1)~,
\label{Pk} 
\eeq 
where $\dt(\k)$ is the Fourier transform of the density contrast
$\delta\rho/\rho$.  Assuming a linear and deterministic galaxy bias,
$\delta(\x) = b_1 \delta_m(\x)$, measurement of the galaxy power
spectrum in redshift surveys in the linear regime gives~\cite{K87}
\beq P(k,\mu)= (1 + 2 \beta \mu^2 + \beta^2 \mu^4) 
\, (b_1\sigma_8)^2 \, P_1(k), 
\label{Plin}
\eeq 
where $\beta \approx \Omega_m^{0.6}/b_1 $, $\mu$ is the cosine of the
angle between the wavector and the line of sight, $\sigma_8$ is the rms
density fluctuation amplitude in spheres of radius $ 8 \Mpc $, and
$P_1(k)$ is the linear mass power spectrum scaled to $\sigma_8=1$.
Equation~(\ref{Plin}) illustrates the degeneracies between the
parameters $\Omega_m$, $b_1$, and $\sigma_8$: measurement of the
quadrupole to monopole ratio of $P(k,\mu)$ via redshift distortions,
for example, yields a measurement of $\beta$, whereas the overall
amplitude of the power spectrum gives in addition an estimate of $b_1
\sigma_8$. Equation~(\ref{Plin}) involves the small angle approximation,
however for the parameter combinations measured here, the results will
be unchanged even if the approximation is dropped.

Additional information is needed to break this degeneracy between
$\Omega_m$, $b_1$, and $\sigma_8$; for example, non-linear effects in
pairwise velocities \cite{FJFDJ99,JFFDJ00} and in density-velocity
comparisons \cite{C99} can be used to constrain different combinations
of $\Omega_m$, $b_1$, and $\sigma_8$.  We use here measurements of the
bispectrum (see also \cite{F94,HBCJ95,MVH97,VHMM98,SCF99}).  The
nonlinearities in the equations of motion of GI induce a characteristic
pattern of three-point and higher order correlations
\cite{F84,SCFFHM98}.  For Gaussian initial conditions, the induced
bispectrum $ B_{123} \equiv B(k_1,k_2,k_3) $,
\beq
\lexp \dt(\k_1) \dt(\k_2) \dt(\k_3) \rexp = 
(2\pi)^3 \dD(\k_1+\k_2+\k_3) \, B_{123} , 
\eeq
scales as $ B(k) \propto [P(k)]^2 $; it is thus useful to examine the
reduced bispectrum amplitude $Q$,
\beq
Q \equiv { B_{123} \over P_1 P_2 + P_1 P_3 + P_2 P_3 } .
\eeq
To leading order in perturbation theory for Gaussian initial conditions,
to an excellent approximation GI predicts that $Q$ is independent of
time and thus of $\sigma_8$ and $\Omega_m$; $Q$ is a function of
triangle shape with a behavior that depends only on the spectral index,
$n_{\rm eff}(k)= d\ln P_1(k)/d\ln k$.

We observe galaxies, not mass.  At scales larger than the characteristic
scale of galaxy formation, it is reasonable to assume that the galaxy
density is given by a local transformation of the underlying mass
density field; that is, for the smoothed density $\delta = b_1 \delta_m
+ \case12 b_2 \delta_m^2 + \ldots$ locally, in which case the galaxy
bispectrum amplitude ($Q$) is related to the matter amplitude ($Q_m$)
\cite{FG93} as
\beq 
Q = {Q_m \over b_1} + {b_2\over b_1^2} . \label{Qbias} 
\eeq 
This shows how the galaxy bispectrum can be used to break the
degeneracies present in linear theory: since $Q_m$ is independent of
$\sigma_8$ and $\Omega_m$, measuring $Q$ for different triangles and
comparing it to the predicted $Q_m$ based upon the measured spectral index
gives separate values for the bias parameters $b_1$ and $b_2$
\cite{F94}.  In redshift space (where distances are measured by
recession velocities using the Hubble law) there are small corrections to
Eq.~(\ref{Qbias}) that can be computed; for $b_1=1$, $b_2=-0.5$, and
$\Omega_m=0.3$, (\ref{Qbias}) underestimates the exact result in the
large-scale limit by $\Delta Q \approx 0.14$, independent of
configuration \cite{SCF99}. Thus, it does not affect $b_1$, and given
the observational uncertainties, it only minimally alters the inferred
$b_2$.

To estimate the galaxy power spectrum and bispectrum, we follow Feldman,
Kaiser, and Peacock \cite{FKP94} in using the optimal weighted transform
of the difference between the data and a tailored synthetic catalog
\cite{PI,PII}.  In addition we construct mock catalogs using
second-order Lagrangian PT (2LPT), which agrees with $N$-body
simulations at the scales we consider \cite{PI}.  The speed of 2LPT
allows us to construct a large number of mock catalogs which take into
account the selection function, survey geometry, and redshift-space
mapping.  This is essential to estimate accurately the mean, errors,
covariance, and the distribution of $Q$ for different triangles,
required to compute the non-Gaussian likelihood function~\cite{PI}.  The
2LPT mock catalog initial conditions are based on a $\Lambda$CDM model
with $\Omega_m=0.3$, $\Omega_\Lambda=0.7$ ($\Lambda$ stands for the
cosmological constant); the linear power spectrum shape parameter
$\Gamma=\Omega_m h =0.21$ is consistent with PSC$z$ power spectrum
measurements \cite{Tetal99,Setal99}, and the adopted normalization,
$\sigma_8 = 0.7$, corresponds to $\sigma_8^s=0.84$ in redshift space,
consistent with the observed normalization for \IRAS galaxies
\cite{FDSYH94}.

\begin{figure}[t!]
\mbox{\epsfxsize=8.6cm \epsffile{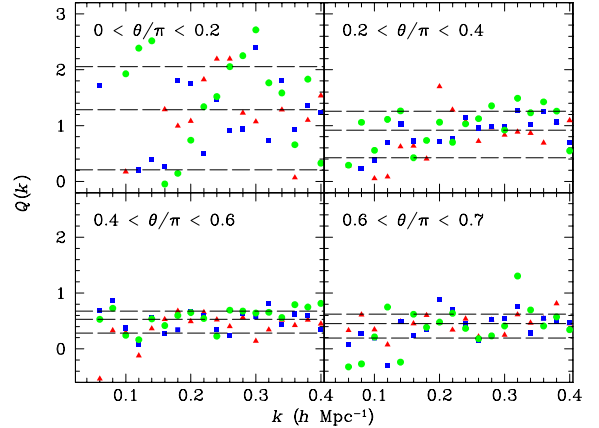}}
\caption {$Q$ from the PSC$z$ survey vs.~$k = k_1$ for triangles
with ratio of sides $k_2/k_1$ separated by angle $\theta$.  Panels show
bands in triangle shape, from nearly collinear (upper left) to open
(lower right).  Triangles, squares, and circles show results for
$k_2/k_1=0.25$--0.5, 0.5--0.75, 0.75--1.0.  The dashed lines in each
panel show the median and 68\% range of the points.
\label{QkPSCz}}
\end{figure}

{\it Data and Analysis.}---The \IRAS PSC$z$ survey~\cite{PSCz} is based
upon the \IRAS Point Source Catalog 
with additions to achieve the best possible uniformity over the sky.  Out of
the 15411 \IRAS galaxies across 84\% of the sky, we take the 13180
galaxies in the range $ 20 \Mpc \le R \le 500\Mpc $, with galactic
latitude $ |b|\ge 10^\circ $, and IRAS 60 micron flux $ f_{60}>0.6 \Jy
$.  Our results appear graphically in Figures 1 and 2.  The bispectrum
is defined for closed triangles in Fourier space.  We characterize a
given triangle by its ordered sides $ k_1 \ge k_2 \ge k_3 $.  In Fig.~1
we show $Q$ from PSC$z$ as a function of $k_1$ for triangles with two
sides separated by angle $\cos \theta = ({\hat \k_1 \cdot \hat \k_2})$,
for four ranges of $\theta$.  Since the width of the survey window
function is $\approx 0.01 \kMpc$, we confine our quantitative analysis
to triangle sides $k_3 \ge 0.05\kMpc$, for which deconvolution using the
narrow window approximation is accurate \cite{PI}.  The median of $Q$
decreases with increasing $\theta$, as predicted by GI \cite{F84} (see
Fig.~2).

Figure 2 shows $Q$ from PSC$z$ as a function of angle $\theta$ for
triangles with $ 0.2 \leq k_1 \leq 0.4 \kMpc$ and with two sides of
ratio $ k_2/k_1 = 0.4 $--0.6.  The solid curve shows the prediction for
$Q$ in redshift space, obtained from averaging many 2LPT mock catalogs
as detailed above.  Symbols show results for bands in $k_1$, as
identified in the caption.  The dependence of $Q$ on triangle shape
predicted by GI is clearly seen in the data.  We employ a likelihood
analysis as in\cite{PI,PII} to quantify the constraints that this places
on galaxy biasing and primordial non-Gaussianity.  For Gaussian initial
conditions, the dashed curve shows the 2LPT $\Lambda$CDM predictions
with the best-fit values $ 1/b_1 = 1.20 $, $ b_2/b_1^2 = -0.42 $, as
given in Table~1.

\begin{figure}[b]
\mbox{\epsfxsize=8.6cm \epsffile{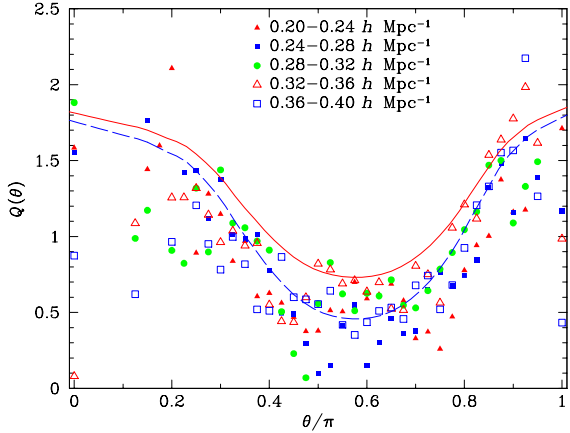}}
\caption {$Q$ from the PSC$z$ survey vs.~$\theta$ for triangles with 
$ 0.2 \leq k_1 \leq 0.4\kMpc$ and with sides of ratio $k_2/k_1=0.4-0.6$
separated by angle $\theta$.  The solid curve shows $Q$ in redshift
space averaged over many 2LPT realizations of the $\Lambda$CDM
model. The dashed curve shows the prediction for $\Lambda$CDM with the
best--fit bias parameters $1/b_1 = 1.20$, $b_2/b_1^2=-0.42$.
\label{QaPSCz}}
\end{figure}

Figure 3 shows the likelihood contours and marginalized distributions
for $ 1/b_1 $ and $ b_2/b_1^2 $ for the PSC$z$ survey, assuming Gaussian
initial conditions and including triangles with $ k < k_{\rm max}=0.3
\kMpc $, where the upper limit is imposed by both nonlinearity and
discreteness.  We have varied the maximum wavenumber allowed in the
analysis, $ k_{\rm max}=0.2 $, 0.25, and $ 0.3 \kMpc $, (see Table~1.)
The maximum likelihood values are essentially independent of $k_{\rm
max}$, though the error bars decrease as $k_{\rm max}$ is increased.
There is no evidence that the bias parameters depend on $k$.  These
results for $b_1$ and $b_2$ are insensitive to the values of $\sigma_8$
and $\Omega_m$ adopted in the mock catalogs
\cite{PI,PII}, but they do depend upon the assumed value for $\Gamma$, 
since it determines the predicted $Q_m$ \cite{PI,PII}. 

If galaxy biasing is local in Lagrangian, rather than Eulerian space
(imposed on the initial, rather than final conditions), then $Q$ in
Eq.(\ref{Qbias}) has an additional term which depends upon configuration
\cite{CPK00}. We have repeated the likelihood analysis for such a model
and found that the likelihood ratio for the Eulerian to Lagrangian
maximum likelihood is ${\cal L}_E/{\cal L}_L =1.6$.  This implies that
the behavior of the bispectrum for PSC$z$ galaxies is better described
by an Eulerian bias scheme.

\begin{table}
\caption{Likelihood Estimates of Bias Parameters}
\begin{center}
\begin{tabular}{crrr}
\sk
parameter & $k_{\max}=0.2$ & $k_\max=0.25$ & $k_\max=0.3$ \\
\sk
\hline 
\sk
$1/b_1$&
$1.15^{+0.36}_{-0.37}$&
$1.11^{+0.22}_{-0.25}$&
$1.20^{+0.18}_{-0.19}$\\
\sk
$b_2/b_1^2$&
$-0.38^{+0.37}_{-0.39}$&
$-0.29^{+0.27}_{-0.24}$&
$-0.42^{+0.19}_{-0.19}$\\ 
\sk
\end{tabular}
\label{comp}
\end{center}
\end{table}

\begin{figure}[t]
\mbox{\epsfxsize=8.6cm \epsffile{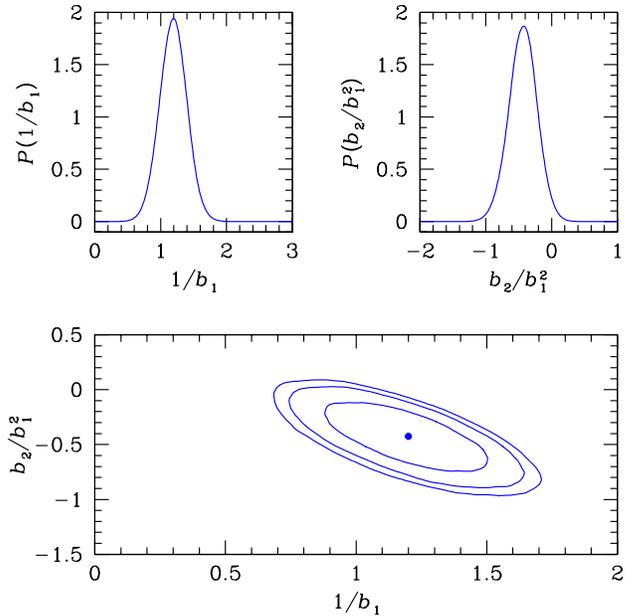}}
\caption {Likelihood contours for bias parameters $b_1$ and $b_2$ 
from the PSC$z$ survey for $k_\max=0.3\kMpc$, 
assuming Gaussian initial conditions and a linear 
power spectrum with $\Gamma=0.21$.
The contours denote $68\%$, $90\%$ and $95\%$ confidence levels.}
\end{figure}

The PSC$z$ bispectrum provides strong constraints upon non-Gaussian
initial conditions.  As an extension of the results in \cite{PII}, we
consider $\chi_N^2$ statistics as a general example of dimensional
scaling models, where the linear PT $n$-point correlation function
$\xi_n \propto \xi_2^{n/2}$.  For $N=1$, this corresponds to the
predictions of some inflationary models with isocurvature perturbations
\cite{P99,LM97}; as $N \to \infty$ the model becomes effectively
Gaussian, and for a fixed power spectrum (taken to fit that of PSC$z$)
the primordial bispectrum obeys $Q^I \propto N^{-1/2}$ \cite{S00}.  In
addition, there are two other contributions that partially cancel that
scale as $N^{-1}$~\cite{S00}.  The amplitude and scaling of the PSC$z$
bispectrum constrains $N$ to be considerably larger than unity.  Thus,
we use $Q(N) \sim Q_G +(Q_{\chi^2}-Q_G) N^{-1/2}$, and similarly for the
errors, where $Q_G$ and $Q_{\chi^2}$ are obtained from mock catalogs for
Gaussian and $N=1$ initial conditions, respectively.  In addition, we
assume that $Q$-eigenmodes and eigenvalues are well approximated by the
Gaussian ones in the large $N$ regime.  We construct a likelihood for
$N$ by marginalizing over the bias; from this we obtain $ N > 49\ (22)$
at 68\% (95\%) CL.  Since the primordial dimensionless skewness is
$B_3=2.46$ for a $\chi_1^2$ field~\cite{P99}, we conclude that the
PSC$z$ bispectrum constrains $ B_3 < 0.35 (0.52)$.

{\it Discussion.}---Our findings are consistent with previous studies
from sparser and smaller \IRAS surveys~\cite{PII}, but using the PSC$z$
catalog reduces the statistical and systematic errors and allows us to
place the strongest constraints to date upon nonlinear bias and
primordial non-Gaussianity.  Our main results are the following:

(1) The dependence of the bispectrum upon configuration shape is the
clearest direct evidence to date that non-linear mode couplings are as
specified in GI.

(2) The linear bias parameter we find, $1/b_1=1.20^{+0.19}_{-0.18}$,
agrees well with $b_1=0.84\pm 0.28$~\cite{TBHT00} obtained by a joint
analysis of the PSC$z$ power spectrum, cosmic microwave background (CMB)
anisotropy, and cluster abundances.  It is also consistent with that
inferred for APM galaxies~\cite{APM} using the shape of the two-point correlation
function, $b_{\rm APM} = 1.11 \pm 0.22$~\cite{GJ00}, taking into account
the relative bias $b_{\rm APM}/b_{\rm PSCz} =1.15$~\cite{HT00}.  It is
in less good agreement but still consistent with $ b_{\rm PSCz} = 1.10
\pm 0.23 $ (95\% CL) found from a concordance between PSC$z$ and recent
CMB results~\cite{TZH00}.

(3) We obtain a nonlinear bias $ b_2/b_1^2=-0.48^{+0.19}_{-0.18} $,
which is not yet measured in any other way.  A negative $ b_2 $ means
that the number of \IRAS galaxies does not rise as quickly as the mass
density contrast; this is consistent with the observation that
infrared-selected galaxies are increasingly under--represented in
massive clusters.

(4) That both the scale and shape dependence of the bispectrum are
consistent with GI acting on Gaussian initial conditions limits the
possibility of primordial non-Gaussianity.  In particular, our study of
$\chi^2_N$ initial conditions constrains $ N > 49 $ and the primordial
skewness to be $ B_3<0.35 $.

(5) Our results for the bispectrum are consistent with the galaxy
skewness $S_3$ measured for the same catalog \cite{SBFMS00} and are
similarly insensitive to the flux cut: we find identical results for
$f_{60} > 0.8\Jy$. Further, the PSC$z$ data is incomplete beyond
$420\Mpc$ \cite{HTP00}, however, since we look at $k>0.05$, this should
not affect the bispectrum signal.

(6) When coupled with measurements of the power spectrum
redshift distortions, which determine $\beta = \Omega^{0.6}_m/b_1 \simeq
0.4 \pm 0.12$ for the PSC$z$ survey \cite{HTP00,TBHT00}, our measurement
of $b_1$ from the bispectrum allows one to break the degeneracy between
linear bias and $\Omega_m$: we find $\Omega_m=0.16 \pm 0.1$, indicating
a low-density universe.  We stress that, unlike other methods, this
result is based only on the clustering properties of PSC$z$ galaxies
assuming Gaussian initial conditions, independent to a large extent of
external determinations of $\sigma_8$. 

Due to its almost complete sky coverage and large volume, the PSC$z$
survey is the best survey available to study the bispectrum at large
scales.  This will soon change with the advent of larger redshift
surveys such as 2dF and SDSS, which will determine the bispectrum even
more precisely and thus test the GI framework, nonlinear galaxy bias,
and primordial non-Gaussianity to unprecedented accuracy.

Research supported in part by the NSF grant AST-0070702 at Kansas, by
the DOE and NASA grant NAG5-7092 at Fermilab, and by NASA NAG5-2835 at
Florida.  RS is supported by endowment funds from the IAS and thanks
M. Zaldarriaga for useful discussions. We thank A. Hamilton and M.
Tegmark for useful comments.

\end{document}